\documentclass[12pt,a4paper]{article}

\usepackage{amssymb}
\usepackage{amsmath, epsfig, color, subfigure,bm}
\usepackage{fullpage}
\usepackage{color}
\usepackage{epsfig}
\usepackage{epstopdf}
\usepackage{subfigure}

\newcommand{\be}{\begin{equation}}
\newcommand{\en}{\end{equation}}

\renewcommand{\vec}[1]{\boldsymbol{#1}}

\begin{document}

\title{Dominant negative Poynting effect \\ in simple shearing of soft tissues}

\author{M. Destrade$^{a}$, C.O. Horgan$^{b}$, J.G. Murphy$^{c,a,\star}$ \\[8pt]
$^{a}$School of Mathematics, Statistics and Applied Mathematics, \\
National University of Ireland Galway, Ireland;  \\[4pt] 
$^{b}$School of Engineering and Applied Science, \\
University of Virginia, Charlottesville, VA 22904, USA; \\[4pt]
$^{c}$Centre for Medical Engineering Research, \\
Dublin City University, Glasnevin, Dublin 9, Ireland.}

\date{$^\star$ corresponding author; email: jeremiah.murphy$@$dcu.ie, \\phone: +353-1-700-8924}

\maketitle

\begin{abstract}

We identify three distinct shearing modes for simple shear deformations of transversely isotropic soft tissue which allow for both positive and negative Poynting effects (that is, they require compressive and tensile lateral normal stresses, respectively, in order to maintain simple shear). The positive Poynting effect is that usually found for isotropic rubber. 
Here, specialisation of the general results to three strain-energy functions that are quadratic in the anisotropic invariants, linear in the isotropic strain invariants and are consistent with the linear theory, suggests that there are two Poynting effects that can accompany the shearing of soft tissue: a dominant \emph{negative} effect in one mode of shear and a relatively small positive effect in the other two modes. 
We propose that the relative inextensibility of the fibres relative to the matrix is the primary mechanism behind this  large negative Poynting effect.

\end{abstract}


%%%%%%%%%%%%%%%%%%%%%%%%%%%

\section{Introduction}

%%%%%%%%%%%%%%%%%%%%%%%%%%%

Shearing deformations of soft tissue have somewhat been neglected in the literature, from an experimental as well as from a modelling point of view, being far less common than the almost ubiquitous tensile and biaxial material characterisation tests.  
However, the large shear of a soft material is a most illuminating testing protocol. Simple shear is achieved by gluing two opposite sides of a cuboid sample to two flat rigid platens, and by displacing one platen parallel to the other \cite{BS}. 
What becomes quickly apparent both in practice and in theory is that this displacement is achieved by applying not only a force in the direction of shear (the direction of motion of the moving platen), but also forces in the direction normal to the platens. 
For isotropic materials, Poynting \cite{Poyn} showed experimentally, and Rivlin \cite{Rivl48} theoretically,  that the normal forces had to be compressive on the platens, as simple shear causes the sample to expand in the direction normal to them. 
Although Poynting \cite{Poyn} demonstrated this normal stress effect for pure torsion, the tendency of a cuboid to expand in simple shear is now widely called the \emph{positive Poynting effect}. 
Recent contributions on the topic include those by Mihai and Goriely \cite{MaG}, Destrade \emph{et al.} \cite{DeMS12} and Horgan and Smayda \cite{HaS}. 

Although it is generally accepted that most existing isotropic materials exhibit the \emph{positive} Poynting effect, the experimental data of Janmey \emph{et al.} \cite{Jan} suggest that biogels reinforced with biological macromolecules such as fibrin display a \emph{negative} Poynting effect, i.e., there is a tendency for the platens to move closer together when subjected to large shears. 
Several different approaches could be adopted to model this unexpected behaviour. For example, one could relax some of the usual conditions imposed on the material response such as the Empirical Inequalities or, alternatively, allow for some degree of compressibility, field inhomogeneity, anisotropy, swelling, etc. The work of Destrade \emph{et al.} \cite{DGMM12}, Horgan and Murphy \cite{HaM1}, Mihai and Goriely \cite{MaG} and Wu and Kirchner \cite{WaK} for example, is illustrative of these different methods.
One could also exploit the microstructure of the reinforced biogels, noting that they are composed of semi-flexible filaments embedded in a soft matrix. 
Although the filaments are distributed isotropically in every direction, they behave differently in traction (strong resistance) than in compression (high compliance) and the contribution of the stretched filaments can dominate the overall response. Thus, in simple shear, the strong pull of the fibres can overcome the weaker push of the compressed fibers in the sheared matrix and therefore bring the platens together.
This avenue of micro-structural modelling was explored by Janmey \emph{et al.} \cite{Jan} and Ogden (private communication).
Here we propose that the same effect can be modelled using the phenomenological theory developed by Spencer \cite{Spencer} for strong fibres embedded in an isotropic matrix. 

Hence we assume homogeneity, incompressibility and \emph{transverse isotropy}, so that the mechanical response of the solid is influenced by a single preferred direction (Section \ref{simple-shear}).
We use material models (introduced by Murphy \cite{me}) which are compatible with linear anisotropic elasticity in the infinitesimal regime.
Such consistency with the linear theory is supported by extensive experimental data, particularly for muscles. The models considered are likely to be good models of the mechanical response of soft tissue in general, given that they are guaranteed to model infinitesimal deformations accurately and that typical physiological strains are only of the order of $10\%$. 

The systematic, comprehensive testing regime for the shearing of soft tissue introduced by Dokos \emph{et al.} \cite{Dokos} is considered in Section \ref{ss}, with three distinct physical modes of shear identified, and we compute the corresponding stress components. 
We show that general considerations of the normal stresses that accompany shearing suggest the existence of both positive and negative Poynting effects, but with the negative effect dominant. 
This large negative Poynting effect occurs only for one of the three modes of shear: it occurs when fibres are originally normal to the platens. This is because the fibres strongly resist the stretch imposed upon them by the shearing deformation and overcome the response of the soft isotropic matrix, with the result that there is a tendency for the platens to come closer together
(preliminary results on materials reinforced with \emph{inextensible} fibres were established by Saccomandi and Beatty \cite{SaBe02}).
This is shown in Section \ref{The Poynting effect}, where we also fit the data of Janmey \emph{et al.} \cite{Jan} to some of our models.

The novelty here then is the prediction of a negative Poynting effect of the same order of magnitude as the shear stress when soft tissue is sheared in the physiological range of strain (as in the experiments of Janmey \emph{et al.} \cite{Jan}) that is explained by a simple physical mechanism (see Figure \ref{shearp}). Although the predictions are only for three special modes of shear for three polynomial models, one can justify this claim by noting that the shearing deformations considered are essentially \emph{canonical shearing modes}, in that every shearing deformation can be considered as a nonlinear superposition of these modes and the strain energies are likely to be representative of the mechanical response of soft tissue, as argued above.

%%%%%%%%%%%%%%%%%%%%%%%%%%%%%%

\section{Simple shear of soft tissues and material models} \label{simple-shear}

%%%%%%%%%%%%%%%%%%%%%%%%%%%%%%

We call $\left(X_1,X_2,X_3\right)$ and $\left(x_1,x_2,x_3\right)$ the Cartesian coordinates of a typical particle in the undeformed and deformed configurations, respectively. 
Then $\vec{F} \equiv \partial \vec{x}/ \partial \vec{X}$ is the deformation gradient tensor (with $J \equiv \det \, \vec{F}$), and  $\vec{B} = \vec{FF}^T$, $\vec{C} = \vec F^T \vec F$ are  the left and right Cauchy-Green deformation tensors, respectively. The corresponding principal \emph{isotropic invariants} are defined by
\be \label{i123}
I_1= \text{tr}(\vec{B}), \quad
I_2=\tfrac{1}{2}\left[I_1^2- \text{tr}\left(\vec{B}^2\right)  \right], \quad
I_3=\det(\vec{B})=J^2.
\en

Consider now a transversely isotropic, non-linearly elastic material with a  preferred direction $\vec{M}$ in the undeformed configuration, physically induced by the presence of parallel fibres embedded in a softer elastic matrix. 
The so-called \emph{anisotropic invariants} are defined as
\begin{equation} \label{inv}
I_4=\vec{M}\vec{.}\vec{CM}, \qquad I_5=\vec{M}\vec{.}\vec{C}^2\vec{M}.
\end{equation}
As is well known, $I_4$ is the square of the stretch experienced by material elements  in the fibre direction: when $I_4 \ge 1$, the fibres are stretched, when $I_4 \le 1$, they are compressed.

As the material is assumed perfectly incompressible,  $I_3 \equiv 1$ and the strain energy density $W$ is therefore a function of only four invariants, i.e., $W=W\left(I_1,I_2,I_4,I_5\right)$. 
The corresponding constitutive law has the form (Ogden \cite{Cism})
\begin{multline} \label{cs}
\vec{\sigma}=-p\vec{I} +2W_1\vec{B}-2W_2\vec{B}^{-1}+2W_4\vec{FM} \otimes \vec{FM}  \\ +
 2W_5\left( \vec{FM} \otimes \vec{BFM}+\vec{BFM} \otimes \vec{FM} \right),
 \end{multline}
where $\vec{\sigma}$ denotes the Cauchy stress, attached subscripts denote partial differentiation of $W$ with respect to the appropriate invariant and $p$ is an arbitrary scalar field. 
To ensure that the stress is identically zero in the undeformed configuration, we require that
\be \label{sf}
2W_1^0-2W_2^0=p^0, \qquad W_4^0+2W_5^0=0,
\en
where the $0$ superscript denotes evaluation in the reference configuration, in which $I_1=I_2=3$, $I_4=I_5=1$. It will also be assumed that the strain-energy vanishes in the undeformed configuration, i.e., that $W^0=0$. 

Merodio and Ogden \cite{MaO3} and Murphy \cite{me} obtained restrictions to ensure the compatibility of the linear and non-linear theories. 
This compatibility requires that
\begin{equation} \label{2}
 2W^0_1 + 2W^0_2 = \mu_T, \qquad 2W_5^0=\mu_L-\mu_T, \qquad
4W^0_{44} + 16W^0_{45} + 16W^0_{55} = E_L + \mu_T-4\mu_L,
\end{equation}
where $\mu_T, \, \mu_L$ are the infinitesimal shear moduli for shearing in planes normal to the fibres and along the fibres, respectively, and $E_L$ is the Young's modulus in the fibre direction (see Vergori \emph{et al.} \cite{Vergo13} for compatibility formulas in orthotropic and monoclinic elasticity).

There is significant experimental evidence, particularly for muscles (see, for example Gennisson \cite{Genn}, Papazoglou \emph{et al.}  \cite{Pap}, Sinkus \emph{et al.} \cite{Sinkus}), to suggest that 
\be \label{mus}
\mu_L>\mu_T,
\en
with an order of magnitude difference recorded in some instances. There is a distinct lack of comprehensive experimental data for soft tissue where biaxial testing is combined with shear testing. One notable recent exception is the work of Morrow \emph{et al.} \cite{Morrow} who showed that 
\be \label{emu}
E_L \gg \mu_L,
\en
for the \emph{extensor digitorum longus} muscles of rabbits. These inequalities between the material constants are assumed to hold in what follows. 

The signs of the partial derivatives of the strain-energy function will play an important role in the following analysis.  
For \emph{isotropic} materials, the so-called Empirical Inequalities, given by
\be \label{ei}
W_1 >0, \qquad W_2 \ge 0,
\en
are often enforced. 
They are classically employed for rubber-like materials to ensure that specific choices for the strain energy function give physically realistic mechanical responses (see Truesdell and Noll \cite{TaN} and Beatty \cite{Beatty} for a discussion). 
In the absence of experimental data to suggest otherwise, they are assumed to hold also for our transversely isotropic materials. 

Now note that strain energies that are additively decomposed into separate isotropic and anisotropic components, i.e. strain energies of the form,
\be \label{sepsef}
W=f\left(I_1,I_2\right)+g\left(I_4,I_5\right),
\en
say, are consistent with the linear theory, in the sense that no restrictions are imposed on the material constants $\mu_T, \, \mu_L, \, E_L$ by assuming such a form. 
For simplicity in what follows, separable forms are assumed. 
Simple polynomial models can be adopted  for transversely isotropic soft tissue, where the strain energies are at least quadratic in $I_4$,  $I_5$ and at least linear in $I_1$, $I_2$. 
Murphy \cite{me} and Feng \emph{et al.} \cite{Bayley} have argued that it is essential that the strain-energy function be a function of both anisotropic invariants when modelling soft tissue and we assume this here, as well as a dependence on the two isotropic invariants. 
These considerations lead to the following simple models of transversely isotropic response:
\begin{align} \label{sef}
& W^\text{I} = \tfrac{1}{2}\mu_T \left[\alpha\left(I_1-3\right) + \left(1-\alpha\right)\left(I_2-3\right)\right]
 \nonumber \\ & \qquad \qquad +\tfrac{1}{2}(\mu_T-\mu_L)\left(2I_4-I_5-1\right) + \tfrac{1}{32}(E_L+\mu_T-4\mu_L)\left(I_5-1\right)^2, \nonumber \\[12pt]
& W^\text{II} = \tfrac{1}{2}\mu_T \left[\alpha\left(I_1-3\right) + \left(1-\alpha\right)\left(I_2-3\right)\right]
 \nonumber \\ & \qquad \qquad +\tfrac{1}{2}(\mu_T-\mu_L)\left(2I_4-I_5-1\right)+ \tfrac{1}{16}(E_L+\mu_T-4\mu_L)\left(I_4-1\right)\left(I_5-1\right), \nonumber \\[16pt]
& W^\text{III}= \tfrac{1}{2}\mu_T \left[\alpha\left(I_1-3\right) + \left(1-\alpha\right)\left(I_2-3\right)\right]
 \nonumber \\ & \qquad \qquad +\tfrac{1}{2}(\mu_T-\mu_L)\left(2I_4-I_5-1\right)+ \tfrac{1}{8}(E_L+\mu_T-4\mu_L)\left(I_4-1\right)^2,
\end{align}
where  $0 < \alpha \le1$ (\cite{me}  first introduced these models with $\alpha = 1$).
The linear dependence on the first two strain invariants in \eqref{sef} reflects the isotropic matrix response of a Mooney-Rivlin model.

The last of these models is a generalisation of the so-called Standard Reinforcing Model
\be \label{SRM}
W=c_1\left(I_1-3\right)+c_2\left(I_4-1\right)^2,
\en
(where $c_1$ and $c_2$ are two positive material parameters),
which is often used in the literature for illustrative purposes of fibre reinforcement. 
We thus call the last model of \eqref{sef}: the Compatible Standard Reinforcing Model. 
In addition to the Standard Reinforcing Model \eqref{SRM} we cite the (non-polynomial) Holzapfel-Gasser-Ogden model,
\be \label{HGO} 
 W = \tfrac{1}{2}\mu (I_1 - 3) + \dfrac{k_1}{2k_2}\left\{\exp\left[k_2\left(I_4 -1\right)^2\right] - 1\right\},
 \en
where $\mu$, $k_1$, $k_2$ are positive material parameters Holzapfel \emph{et al.} \cite{HGO}, which is very popular for modelling biological soft tissues.
These two models have a strain energy such that the strain energy depends on only $I_1,I_4$.
We show in Section \ref{The Poynting effect} that a dependence of $W$ on $I_2$ is crucial to capturing the Poynting effect. 
In anticipation, we now introduce a modification of these models to include a Mooney-Rivlin response for the isotropic matrix:
\begin{align}\label{sef45}
& W^\text{IV}= c_1\left[\alpha\left(I_1-3\right) + \left(1-\alpha\right)\left(I_2-3\right)\right] + c_2\left(I_4-1\right)^2, \nonumber \\
& W^\text{V}= \tfrac{1}{2}\mu \left[\alpha\left(I_1-3\right) + \left(1-\alpha\right)\left(I_2-3\right)\right] + \dfrac{k_1}{2k_2}\left\{\exp\left[k_2\left(I_4 -1\right)^2\right] - 1\right\},
 \end{align}
where  $0 < \alpha < 1$.
The first of these was introduced by Le Tallec \cite{LeTa94} and  was used by Horgan and Murphy \cite{HaM1} to demonstrate positive and negative Poynting effects in simple shear; the second one is used in the Finite Element code 
ADINA \cite{adina}.

%%%%%%%%%%%%%%%%%%%%%%%

\section{Stress components in three shear modes} \label{ss}

%%%%%%%%%%%%%%%%%%%%%%%%

\begin{figure} [htp!]
\begin{center}
\emph{(a)} 
\subfigure{\epsfig{width=0.13\textwidth, figure=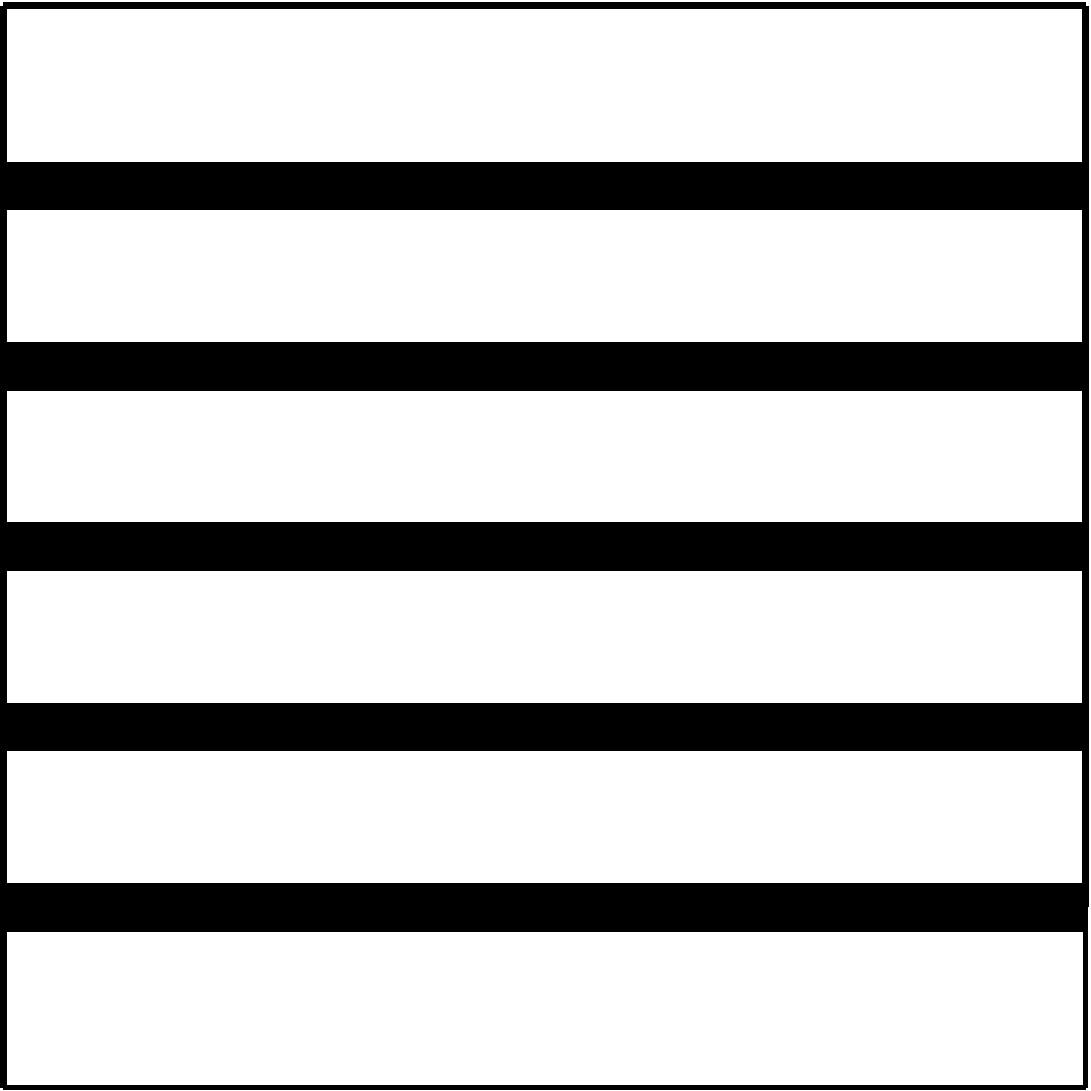}}\qquad
\subfigure{\epsfig{width=0.22\textwidth, figure=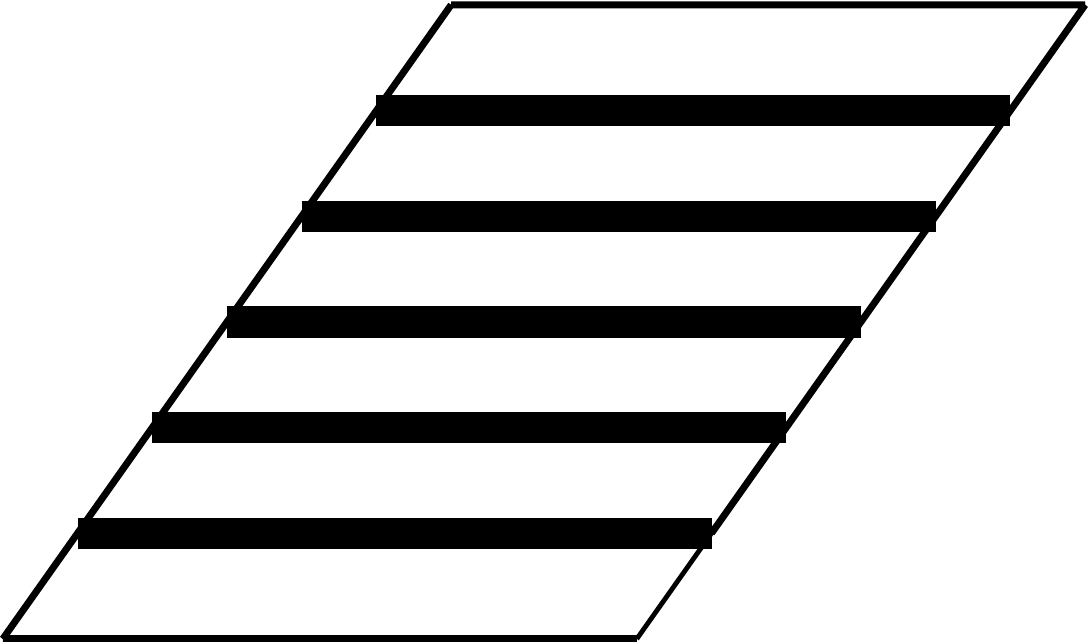}}\qquad
\emph{(b)}
\subfigure{\epsfig{width=0.13\textwidth, figure=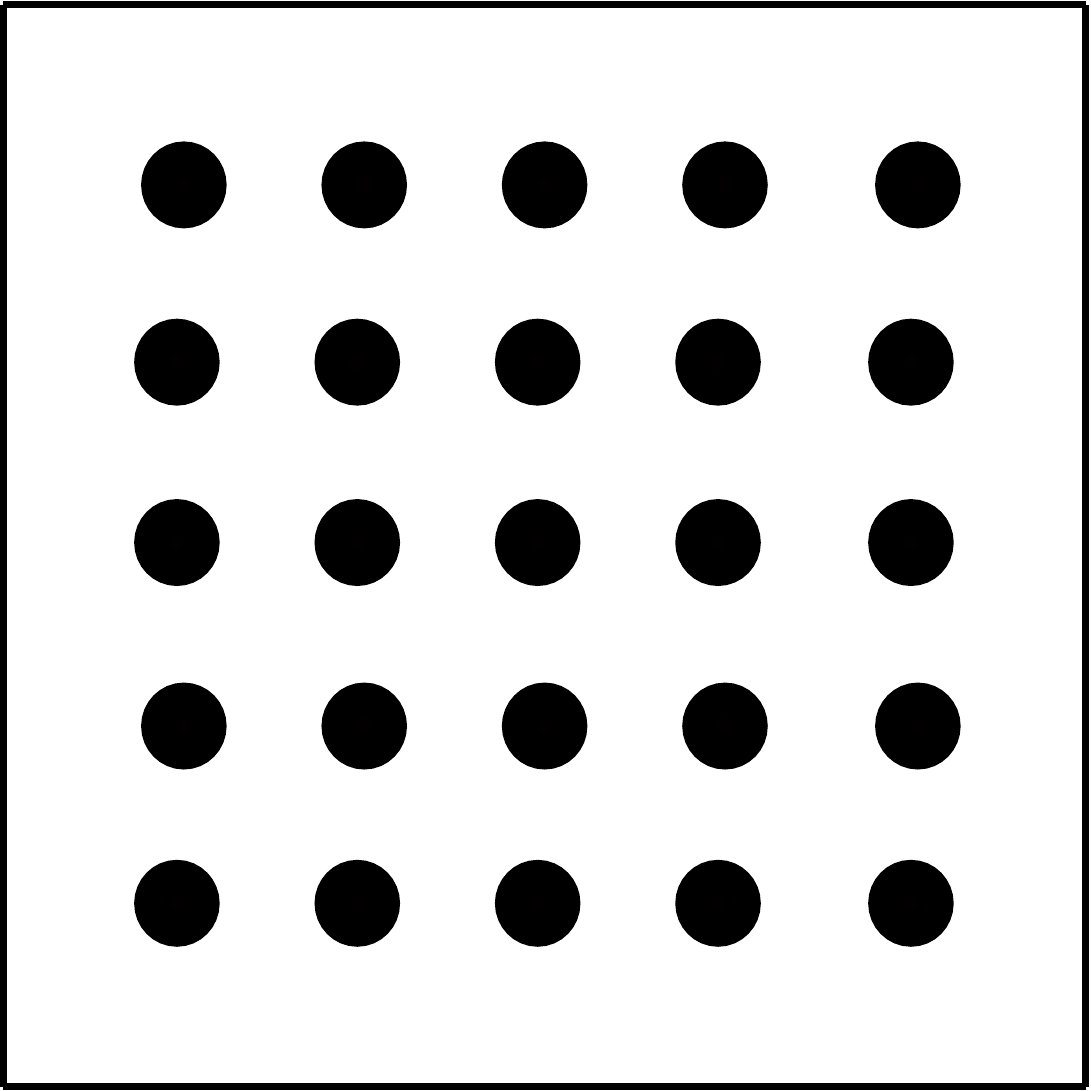}}\qquad
\subfigure{\epsfig{width=0.22\textwidth, figure=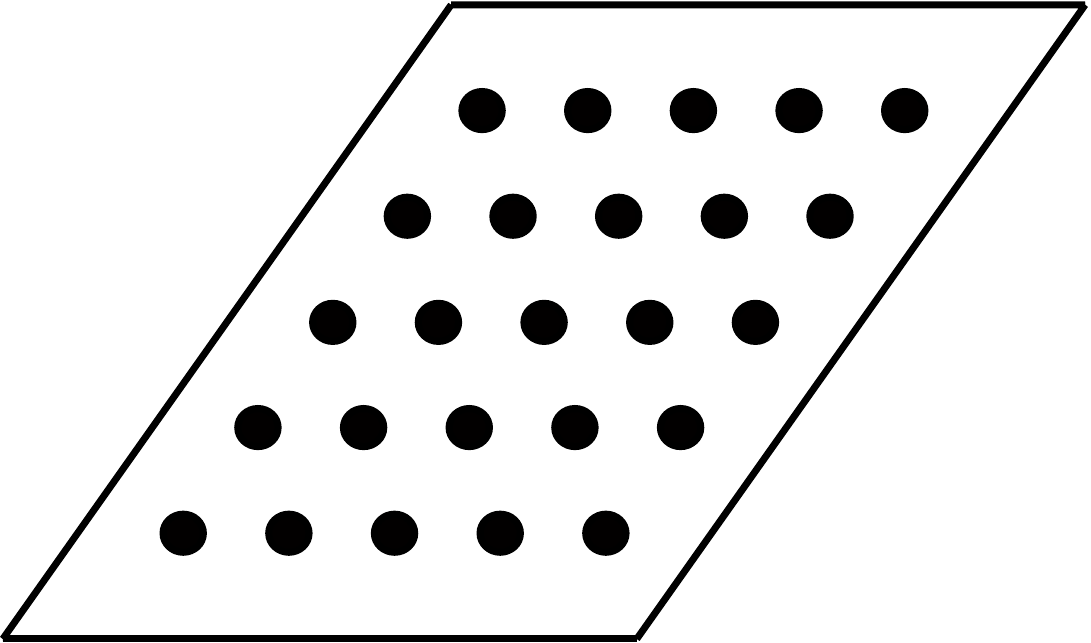}}\\[12pt]
\emph{(c)}
\subfigure{\epsfig{width=0.18\textwidth, figure=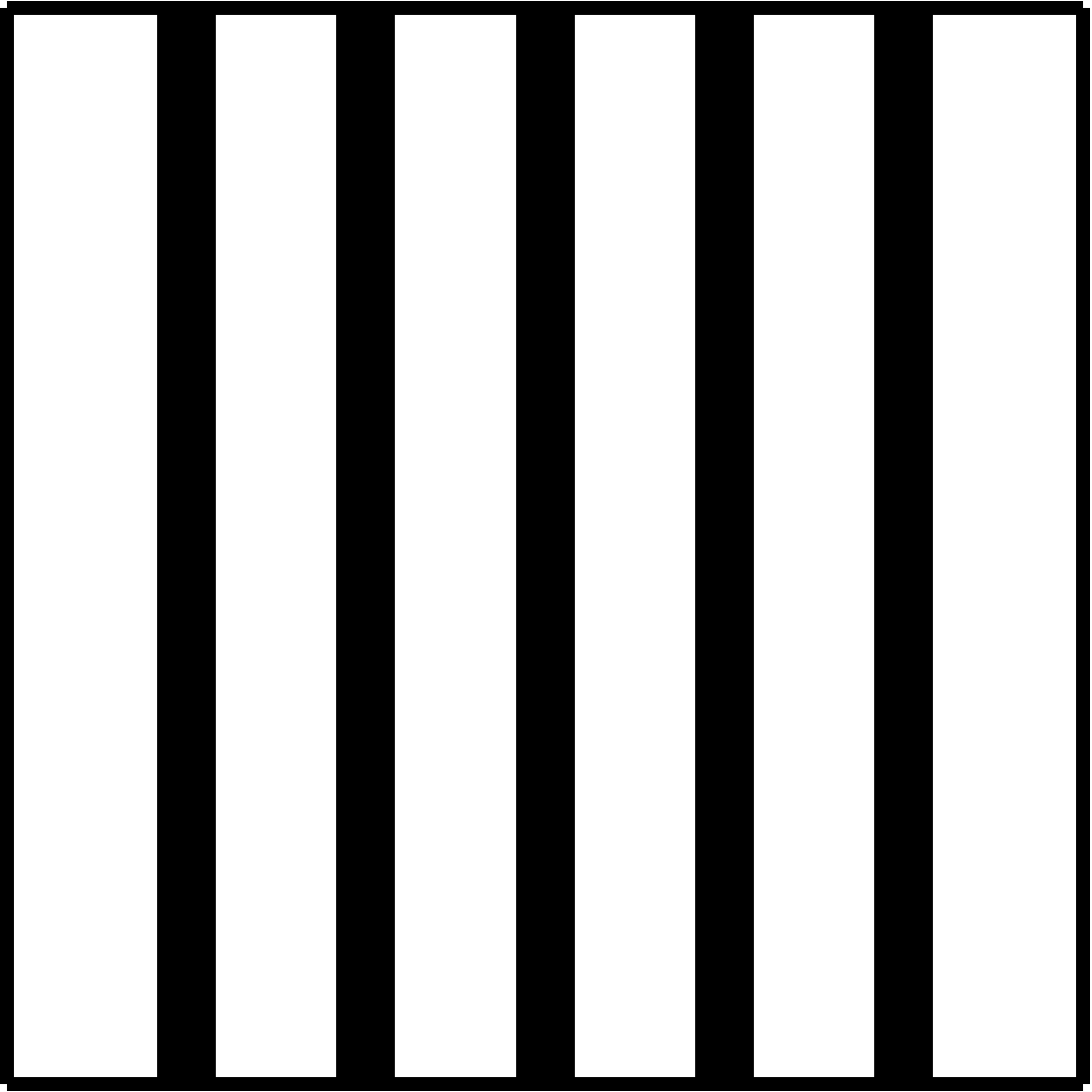}}\qquad
\subfigure{\epsfig{width=0.29\textwidth, figure=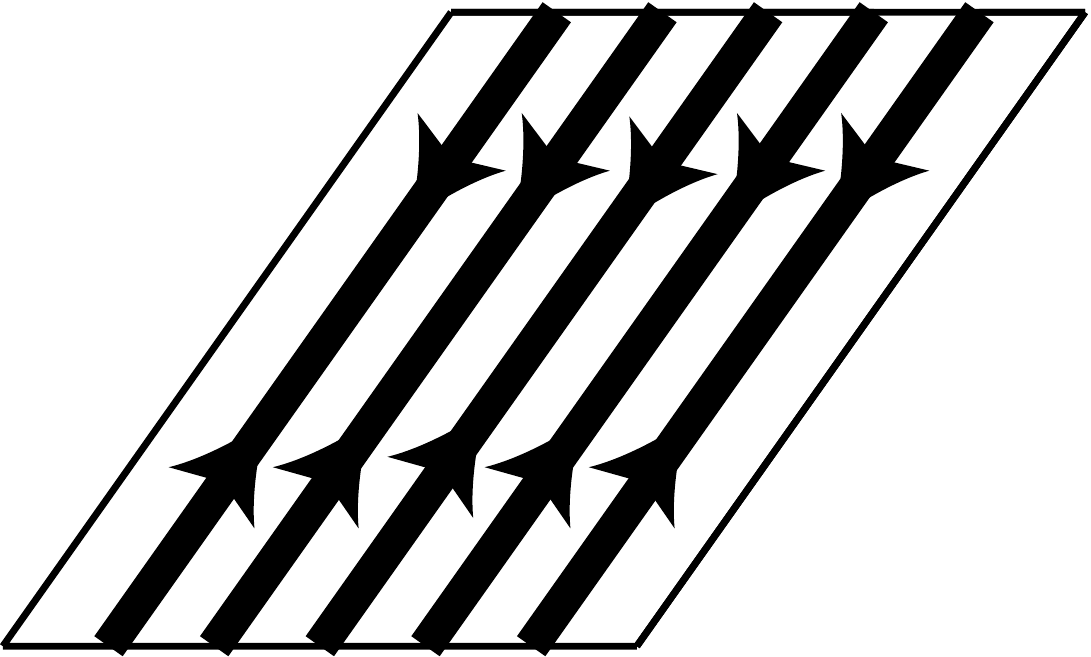}}
\end{center}
\caption{{\small Schematics of \emph{(a)}  Longitudinal Shear; \emph{(b)}  Transverse Shear; \emph{(c)} Perpendicular Shear. In \emph{(a)} and \emph{(c)}, the dark lines represent the fibres. In \emph{(b)}, the darkened circles  denote fibres out of the plane of the page. The arrows in \emph{(c)} indicate the resistive forces developing in the stretched fibres in response to the simple shear: this shear mode provides a simple mechanism to model negative Poynting effects.}}
\label{shearp}
\end{figure}

To investigate the non-linear shear response of passive ventricular myocardium, Dokos \emph{et al.} \cite{Dokos} used cuboids of fibre-reinforced material with a family of parallel fibres aligned with two opposite parallel faces of the block. 
As sketched in Figure \ref{shearp}, there are 3 distinct physical shear responses: \emph{(a)} shearing in the direction of the fibres, which we call Longitudinal Shear; \emph{(b)} shearing in the planes normal to the fibres, which we call Transverse Shear and \emph{(c)} shearing across the fibres, which we call Perpendicular Shear.

The invariants and Cauchy stress components for each of these deformations are given next using the Cartesian representations for the deformations given in \cite{me}, where it was assumed that the $Z-$axis was aligned in the direction of the fibres in the reference configuration. In each case the normal stress to the plane of shear is assumed identically zero.

%++++++++++++++++++++++++++

\subsection{Longitudinal shear}

%++++++++++++++++++++++++++

This shearing mode can be described by the deformation
\be \label{lsd}
x=X, \quad y=Y, \quad z=Z+\kappa X,
\en
giving 
\be
I_1=I_2=3+\kappa^2, \qquad I_4=1, \qquad I_5=1+\kappa^2.
\en
Note that the fibres are not stretched in this shearing mode. 
The corresponding non-zero Cauchy stress components are then obtained from \eqref{cs}  as
\begin{align}
& \sigma_{xx} = -p+2W_1-2W_2\left(1+\kappa^2\right), \nonumber \\
& \sigma_{yy} =-p+2W_1-2W_2, \nonumber \\
& \sigma_{xz}=2\kappa\left(W_1+W_2+W_5\right),\nonumber \\
& \sigma_{zz}=-p+2W_1\left(1+\kappa^2\right)-2W_2+2W_4+4W_5\left(1+\kappa^2\right).
\end{align}
Assuming plane stress conditions, i.e., assuming that $\sigma_{yy}\equiv 0$, determines $p$ as $p=2W_1-2W_2$. 
The remaining normal stress components are therefore
\be
\sigma_{xx}=-2W_2\kappa^2, \qquad
\sigma_{zz} = 2W_1\kappa^2+2W_4+4W_5\left(1+\kappa^2\right).
\en
The normal stress component $\sigma_{xx}$ is the force per current unit area that must be exerted on the $x-$planes in order to maintain the state of simple shear described in \eqref{lsd}. Its sign is determined by the sign of $W_2$. 
The shear stress component $\sigma_{xz}$ is the force per unit area that must be applied in the $x-$direction.

%++++++++++++++++++++++++++

\subsection{Transverse shear}

%++++++++++++++++++++++++++

This deformation can be described as follows:
\be \label{tsd}
x=X+\kappa Y, \quad y=Y, \quad z=Z,
\en
so that
\be
 I_1=I_2=3+\kappa^2, \qquad I_4=1, \qquad I_5=1.
 \en
Again, there is no stretch occurring in the fibre direction.
The Cauchy stress components are 
\begin{align}
& \sigma_{xx} = -p + 2W_1\left(1+\kappa^2\right)-2W_2, \nonumber \\
& \sigma_{xy} = 2\kappa\left(W_1+W_2\right), \notag \\
& \sigma_{yy} = -p+2W_1-2W_2\left(1+\kappa^2\right), \nonumber \\
& \sigma_{zz} = -p+2W_1-2W_2+2W_4+4W_5.
\end{align}
Assuming plane stress conditions means that $\sigma_{zz} \equiv 0$ and therefore that
\be
\sigma_{xx} = 2W_1\kappa^2-2W_4-4W_5, \qquad
\sigma_{yy} = -2W_2\kappa^2-2W_4-4W_5.
\en
Here, the $\sigma_{yy}$ term is the force per unit area that needs to be applied in the direction normal to the shearing platens in order to maintain simple shear, while $\sigma_{xy}$ is applied in the direction of shear. 

%++++++++++++++++++++++++++

\subsection{Perpendicular shear}

%++++++++++++++++++++++++++

This mode of shear can be described by deformations of the form
\be \label{psd}
x = X +\kappa Z, \quad y=Y, \quad z=Z,
\en
giving 
\be
 I_1=I_2=3+\kappa^2, \qquad I_4=1+\kappa^2, \qquad I_5=\left(1+\kappa^2\right)^2+\kappa^2.
 \en
Note that perpendicular shear is the only mode of simple shear in which the fibres are stretched ($I_4 > 1$). Since the fibres are much stiffer than the matrix in which they are embedded, Figure 1(c) suggests that a \emph{tensile} normal force must be applied to the upper and lower surfaces of the specimen to simultaneously stretch the fibres and maintain the specimen in a state of simple shear. Tensile normal forces are equivalent to a negative Poynting effect and perpendicular shear is proposed here as a simple explanation for the mechanism underlying the negative Poynting effect seen in the experiments of Janmey \emph{et al.} \cite{Jan}. This is discussed further in Section \ref{The Poynting effect}.

The non-zero Cauchy stress components for this deformation are 
\begin{align}
& \sigma_{xx}=-p+2W_1\left(1+\kappa^2\right)-2W_2+2W_4\kappa^2+4W_5\kappa^2\left(2+\kappa^2\right), \nonumber \\
& \sigma_{yy}=-p+2W_1-2W_2, \nonumber \\
& \sigma_{xz}=2\kappa\left[W_1+W_2+W_4 +W_5\left(3+2\kappa^2\right)\right], \nonumber \\
&\sigma_{zz}=-p+2W_1-2W_2\left(1+\kappa^2\right)+2W_4+4W_5\left(1+\kappa^2\right).
\end{align}
Setting $\sigma_{yy} \equiv 0$ yields $p$ and therefore
\be
\sigma_{xx}=2\kappa^2\left[W_1+W_4 +2W_5\left(2+\kappa^2\right)\right],  \qquad
\sigma_{zz}=-2W_2\kappa^2+2W_4+4W_5\left(1+\kappa^2\right).
\en
In this case it is the $\sigma_{zz}$ term that determines the force that needs to be applied in the direction normal to the shearing platens in order to maintain simple shear, while $\sigma_{xz}$ is applied in the direction of shear.

%%%%%%%%%%%%%%%%%%%%%%%%

\section{The Poynting effect}
\label{The Poynting effect}

%%%%%%%%%%%%%%%%%%%%%%%%

The normal forces needed to maintain simple shear for each of the three shearing modes derived in the previous section are of interest here. Call the necessary normal stresses for longitudinal, transverse and perpendicular shear $\mathcal{N}_L$,  $\mathcal{N}_T$ and $\mathcal{N}_P$, respectively. Collecting the results of the last section together, these are therefore given by 
\begin{align} \label{ns}
& \mathcal{N}_L = -2W_2\kappa^2, \nonumber \\
& \mathcal{N}_T = -2W_2\kappa^2-2W_4-4W_5, \notag \\
& \mathcal{N}_P = -2W_2\kappa^2+2W_4+4W_5\left(1+\kappa^2\right).
\end{align}

Many models of transversely isotropic soft tissue assume strain-energy functions independent of $I_2$ for simplicity. However, as we can see from \eqref{ns}, these models implicitly assume a Poynting effect that is identically zero in longitudinal shear, a prediction that seems unduly prescriptive and for which there is no experimental justification. 
Hence, in particular, the Standard Reinforcing model \eqref{SRM} and the Holzapfel-Gasser-Ogden model \eqref{HGO} cannot account for a Poynting effect in longitudinal shear.
If, as is almost certainly the case in practice, $W_2>0$, then a compressive force must be applied or otherwise the material would expand normal to the direction of shear. 
This corresponds to the usual \emph{positive Poynting effect}. 
For \emph{isotropic} materials, the crucial role of this dependence on the second invariant was pointed out explicitly by Horgan and Murphy \cite{HaM1}, Horgan and Smayda \cite{HaS} and Mihai and Goriely \cite{MaG}.

Next we note that for separable strain energy functions \eqref{sepsef} in transverse shear (for which $I_4=I_5=1$), we have
\be
2W_4+4W_5=2g_4(1,1)+4g_5(1,1)=0,
\en
because of the initial condition \eqref{sf}$_2$. On using this condition in the second of \eqref{ns}, we see that there are just two modes of normal stress response, the effectively isotropic response
\be \label{iso}
\mathcal{N}_\text{iso} = \mathcal{N}_L=\mathcal{N}_T = -2W_2\kappa^2,
\en
and the perpendicular shear anisotropic response $\mathcal{N}_P$. 

By virtue of the Empirical Inequalities \eqref{ei}, $\mathcal{N}_\text{iso}\le0$, so that we have the usual positive Poynting effect for materials with additively split strain energies in longitudinal and transverse shear, if the strict inequality sign holds.  
Since the anisotropic invariants are associated with the stiff reinforcing fibres, as reflected in the linearisation condition \eqref{2}$_3$ for example, it also seems reasonable to assume that 
\be
W_4 \: \text{ or } \; W_5 \gg W_1, W_2.
\en
If these constitutive inequalities hold then it follows that $\mathcal{N}_P>0$, so that we have a possible \emph{negative Poynting effect in Perpendicular Shear}.

Thus both positive and negative Poynting effects are likely for soft tissue reinforced with macromolecular fibrils and  the magnitude of the negative effect is expected to be much larger. 
These predictions are supported by consideration of specific strain energies, as shown next. 

First note that the isotropic response is the same for all three materials of the polynomial form $W^\text{I}-W^\text{III}$ in \eqref{sef} and is given by 
\be \label{Ni}
\mathcal{N}_\text{iso} =  \mu_T\left(\alpha - 1\right)\kappa^2<0.
\en
Thus a compressive force must be applied to top and bottom surfaces of the sheared specimen to counteract the tendency of these materials to expand in the direction normal to the direction of the applied shear force. 
Thus the classical (positive) Poynting effect occurs for both transverse and longitudinal shears. 

\begin{figure} [htp!]
\begin{center} \epsfig{width=0.6\textwidth, figure=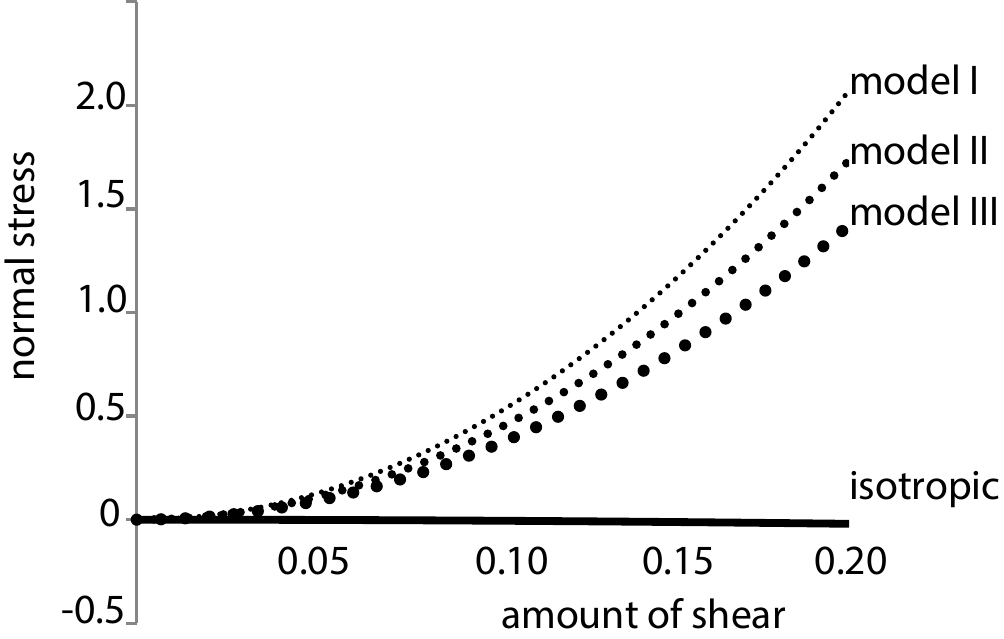}\end{center}
\caption{{\small The normal stress responses accompanying perpendicular shear (dashed lines) and the isotropic transverse and longitudinal shear (full line) for three models of solids reinforced with one family of fibres. }}
\label{com}
\end{figure}

Using an obvious notation, the anisotropic normal response in perpendicular shear of the three models is given by 
\begin{align} \label{N-I-III}
& \mathcal{N}_P^\text I = \left[ \mu_T\left(\alpha - 3\right)+2\mu_L+\tfrac{1}{4}(E_L+\mu_T-4\mu_L)\left(1+\kappa^2\right)\left(3+\kappa^2\right)\right]\kappa^2, \nonumber \\[8pt]
& \mathcal{N}_P^\text{II} = \left[ \mu_T\left(\alpha - 3\right)+2\mu_L + \tfrac{1}{8}(E_L+\mu_T-4\mu_L)\left(5+3\kappa^2\right)\right]\kappa^2, \nonumber \\[8pt]
&\mathcal{N}_P^\text{III} = \left[ \mu_T\left(\alpha - \tfrac{5}{2}\right)+\tfrac{1}{2}E_L\right]\kappa^2. 
\end{align}
Now let $\mathcal{N}/\mu_T$ be a dimensionless measure of the normal stress. 
For a physiological range of strain, each of the normalised normal stresses for perpendicular shear is plotted in Figure \ref{com} for 
\begin{itemize}
\item $\alpha = 1/2$, corresponding to middle of the range of the parameter;
\item $\mu_L/\mu_T=5$, corresponding to a typical relationship between the shear moduli for the experimental data given in Gennisson \cite{Genn}, Papazoglou \emph{et al.}  \cite{Pap}, Sinkus \emph{et al.} \cite{Sinkus};
\item $E_L/\mu_T=75$, motivated by the data of Morrow \emph{et al.} \cite{Morrow}.
\end{itemize}
The normalised isotropic response \eqref{Ni} is also plotted in Figure \ref{com} for $\alpha = 1/2$.

Recall that a positive (tensile) normal stress corresponds to the \emph{negative} or reverse Poynting effect.
From Figure \ref{com}, we see that  perpendicular shear is accompanied by a \emph{negative} Poynting effect and that relative to the perpendicular shear response, the Poynting effect accompanying transverse and longitudinal shear is approximately zero (although in absolute terms, there is a (small) positive Poynting effect). 
Therefore the magnitude of the negative Poynting effect dominates the positive effect when soft tissue is sheared. 
From Figure \ref{com}, it is also seen that the tensile normal stress is monotonically increasing with respect to the amount of shear. 

\begin{figure} [htp!]
\begin{center} \epsfig{width=0.6\textwidth, figure=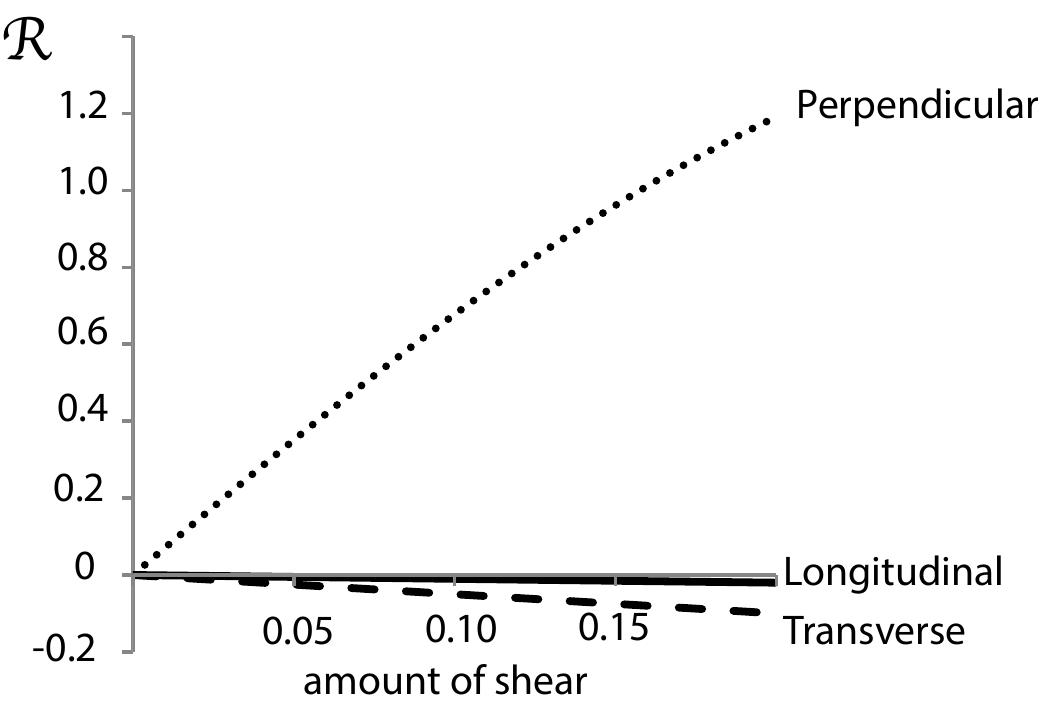}\end{center}
\caption{{\small Plots of $\mathcal{R}$ (relative magnitude of normal stress with respect to applied shear stress) for the different shearing modes of one model of solid reinforced with one family of fibres.}}
\label{model2}
\end{figure}

The relative magnitudes of the normal stresses with respect to the applied shear stresses are now investigated. 
Define the dimensionless quantity 
\be
\mathcal{R} \equiv \frac{\sigma_\text{normal}}{\sigma_\text{shear}}.
\en
For simplicity, we only consider the so-called Compatible Standard Reinforcing Model (Model III, given by \eqref{sef}$_3$), since it gives rise to the smallest normal stress response for perpendicular shear (see Figure \ref{com} above). 
The corresponding longitudinal, transverse and perpendicular shear stresses are 
\be
\sigma_\text{shear}^L = \mu_L\kappa,      \qquad 
\sigma_\text{shear}^T =  \mu_T\kappa, \qquad
\sigma_\text{shear}^P =  \mu_L\kappa +\tfrac{1}{2}(E_L-3\mu_T)\kappa^3,
\en
respectively.
The corresponding $\mathcal{R}$ for each mode of shear is therefore 
\be
\mathcal{R}^L =\frac{\mu_T}{\mu_L}\left(\alpha-1\right)\kappa,      \qquad 
\mathcal{R}^T =  \left(\alpha-1\right)\kappa,       \qquad
\mathcal{R}^P =  \frac{ \left[ \mu_T\left(2\alpha - 5\right)+E_L\right]\kappa}{ 2\mu_L +\left(E_L-3\mu_T\right)\kappa^2 },
\en
respectively. 
For the material parameters used previously, these ratios are plotted in Figure \ref{model2}.

 There, even for the restricted amount of shear considered (in line with physiological strains being of the order of $10\%$), a physically significant reverse Poynting effect is shown to accompany perpendicular shearing deformations, with the corresponding effect for the other two modes of shear relatively unimportant. 
 Indeed for the upper range of shear considered, it is seen that the tensile normal stress in perpendicular shear is of the same order of magnitude as the shearing stress. 
 While the existence of a substantial Poynting effect in soft tissue awaits experimental confirmation, the data of Janmey \emph{et al.} \cite{Jan} for semi-flexible biopolymer gels confirm the existence of a Poynting effect of this order of magnitude. 
 This suggests that the Poynting effect could be an important physiological control mechanism, for example.

\begin{figure} [htp!]
\begin{center} \epsfig{width=0.45\textwidth, figure=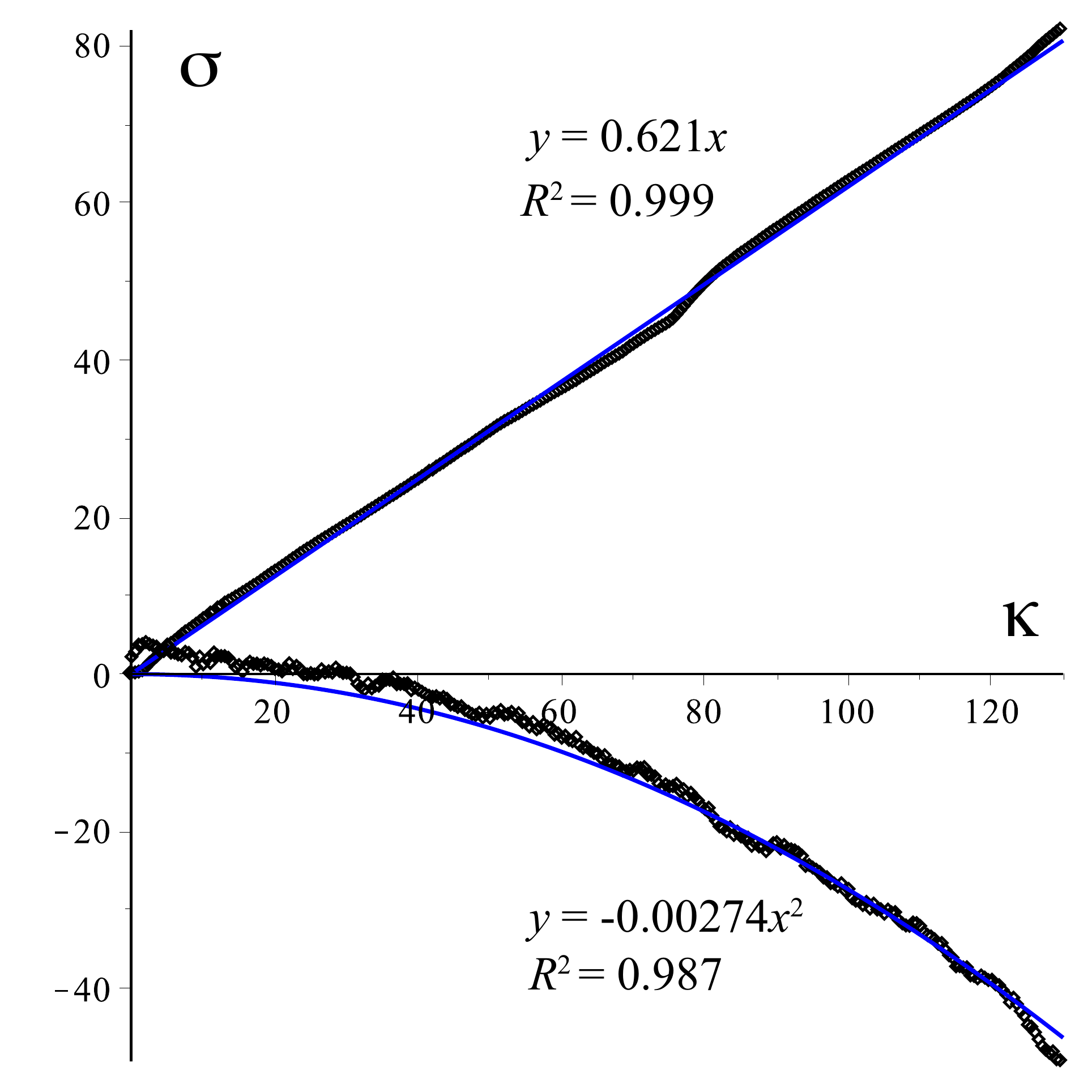}\end{center}
\caption{{\small Data of Janmey \emph{et al.} \cite{Jan} for the shearing of  a block of  gel made from actin cross-linked by polyacrylamide, displaying a linear shear stress-amount of shear relationship and a quadratic normal stress-amount of shear  relationship. More than 260 measurements were recorded.
The negative sign of the normal stress indicates the classical (positive) Poynting effect.
The actual values are unimportant, as both the stress and strain are computed up to multiplicative constants (see for instance the ARES rheometer manual\cite{manual}). }}
\label{fig:PA}
\end{figure}

Finally, we analyse two of the data sets presented by Janmey \emph{et al.} \cite{Jan}: one exhibiting the positive Poynting effect, the other, the reverse Poynting effect. When shearing a block of  gel made from actin cross-linked by polyacrylamide, Janmey \emph{et al.} \cite{Jan} observed a positive Poynting effect. 
In Figure \ref{fig:PA} we see that the fitting of the shear and normal stress components to linear and quadratic trends, respectively, gives excellent results ($R^2=0.999, 0.987$, respectively), i.e. that the material constitutive law obeys
\begin{equation}\label{PA}
\sigma_\text{shear} = a \kappa, \qquad \sigma_\text{normal} = - b \kappa^2,
\end{equation}
for some positive constants $a$ and $b$. 
If the polyacrylamide gel is an isotropic material, then \eqref{PA} clearly indicates that it could be modelled as a Mooney-Rivlin material.
If it is a transversely isotropic material, and biaxial testing is the most efficient way of determining this, then the data reported in the figure cannot correspond to perpendicular shear (which would not give a positive Poynting effect). 
However, polyacrylamide could be modelled with the strain energies $W^\text{I}$--$W^\text{V}$  of \eqref{sef} and \eqref{sef45} as subject to transverse shear, because all models predict $\sigma_\text{shear}$ and $\sigma_\text{normal}$ of the form \eqref{PA}.
It could also have been subject to longitudinal shear because all models predict $\sigma_\text{shear}$ and $\sigma_\text{normal}$ of the form \eqref{PA}, except for Model $W^\text{I}$, which gives $\sigma_\text{shear}^\text{I}$ as an odd cubic in $\kappa$.
 
\begin{figure} [htp!]
\begin{center} \epsfig{width=0.45\textwidth, figure=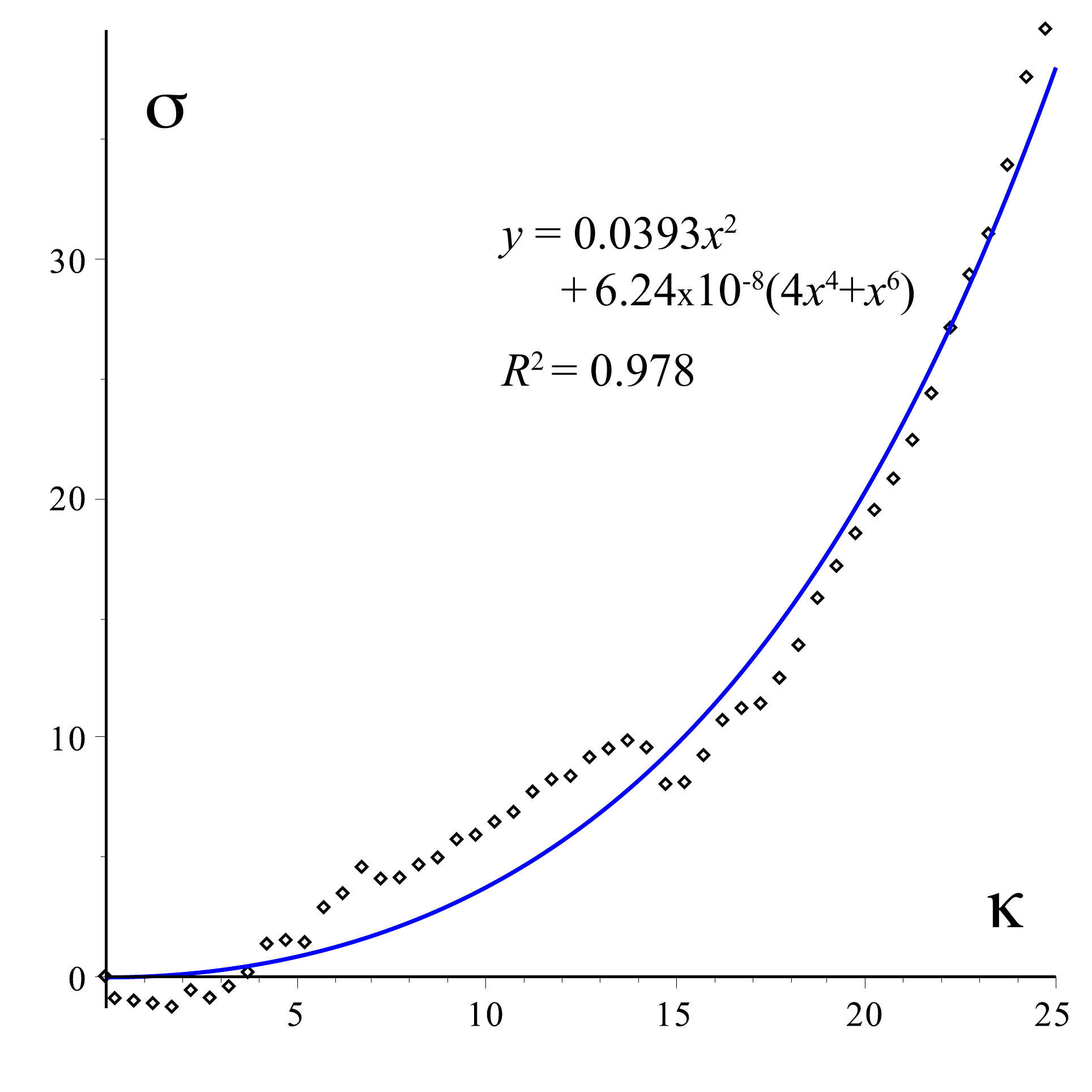}\end{center}
\caption{{\small Normal stress vs. amount of shear from the data of Janmey \emph{et al.} \cite{Jan} for the shearing of  a block of  gel made from actin cross-linked by collagen.
About 50 measurements were recorded.
The (mostly) positive sign of the normal stress indicates a reverse (negative) Poynting effect.
The actual values are unimportant, as both the stress and strain are computed up to multiplicative constants (see for instance \cite{manual}). The experimental data are fitted with the model \eqref{N-I-III}$_1$ in perpendicular shear. }}
\label{fig:collagen}
\end{figure}

\vspace{12 cm}
In Figure \ref{fig:collagen} we present the data collected by Janmey \emph{et al.} \cite{Jan}, showing a reverse Poynting effect for the shear of a block of gel crossed-linked with collagen. 
This phenomenon cannot be captured by isotropic models (unless the empirical inequalities \eqref{ei} are violated).
For transverse isotropy, we have perpendicular shear at our disposal to model the data. 
We used in turn the five models  $W^\text{I}$--$W^\text{V}$  of \eqref{sef} and \eqref{sef45} and found that $W^\text{I}$ provided the best fit ($R^2=0.978$), see the figure.
We used \eqref{N-I-III}$_1$ as an objective function, writing $\mathcal{N}_P^\text{I}$ as $\mathcal{N}_P^\text{I} = a \kappa^2 + b(4\kappa^4+\kappa^6)$ where $a$ and $b$ are best-fit parameters to be determined. 
Of course, we acknowledge that the fitting exercise has its limitations, given that the data is not smooth and we have no way of knowing whether this particular gel was both transversely isotropic and subjected to perpendicular shear.

\vspace{6 cm}

%%%%%%%%%%%%%%%%%%%%%%%%%

\section{Conclusion}

%%%%%%%%%%%%%%%%%%%%%%%%%

The simple analysis presented here, which assumes that plane stress conditions hold when shearing, suggests that for soft tissue a negative (reverse) Poynting effect should occur for perpendicular shear, with the positive effect accompanying the other two modes relatively unimportant. 
It has been demonstrated that the negative Poynting effect could be quite substantial, of the same order of magnitude of the applied shearing stress for physiological strains in some circumstances. 
While these results are consistent with experimental data of Janmey \emph{et al.} \cite{Jan} for hydrogels, further experimental work on biotissues is clearly desirable.

%%%%%%%%%%%%%%%%%%%%%%%%

\section*{Acknowledgements}

%%%%%%%%%%%%%%%%%%%%%%%%

We thank Professor Paul Janmey for providing us with the data of the experiments published in Janmey \emph{et al.} \cite{Jan}.
JGM would like to thank Professor Alain Goriely for stimulating discussions on this and other topics. The constructive criticism of the anonymous referees is also gratefully acknowledged.

%%%%%%%%%%%%%%%%

%\bibliographystyle{elsarticle-harv}


\begin{thebibliography}{00}

\bibitem{BS}
British Standard BS ISO 8013:2006
Rubber, vulcanized --- Determination of creep in compression or shear.

\bibitem{Poyn}
Poynting JH (1909) 
On pressure perpendicular to the shear planes in finite pure shears, and on the lengthening of loaded wires when twisted.
Proceedings of the Royal Society of London. Series A 82:546--559.

\bibitem{Rivl48}
Rivlin RS (1948)
Large elastic deformation of isotropic materials IV: Further developments of the general theory.
Philosophical Transactions of the Royal Society of London. Series  A 241:379--397.

\bibitem{MaG}
Mihai LA, Goriely A (2011) 
Positive or negative Poynting effect? The role of adscititious inequalities in hyperelastic materials. 
Proc. Roy. Soc. Lond. A 467:3633--3646.

\bibitem{DeMS12}
Destrade M, Murphy JG, Saccomandi G (2012) 
Simple shear is not so simple, 
International Journal of Non-Linear Mechanics 
47:210--214.

\bibitem{HaS}
Horgan CO, Smayda M (2012) 
The importance of the second strain invariant in the constitutive modeling of elastomers and soft biomaterials. 
Mech. Mat. 51:43--52.

\bibitem{Jan}
Janmey PA, McCormick ME, Rammensee S, Leight JL, Georges PC, MacKintosh FC (2007) 
Negative normal stress in semiflexible biopolymer gels. 
Nature Materials 6:48--51.

\bibitem{DGMM12}
 Destrade M, Gilchrist MD, Motherway J, Murphy JG (2012) 
Slight compressibility and sensitivity to changes in Poisson's ratio. 
International Journal for Numerical Methods in Engineering 
90:403--411.
 

\bibitem{HaM1}
Horgan CO, Murphy JG (2011)
On the normal stresses in simple shearing of fiber-reinforced nonlinearly elastic materials. 
J. Elasticity 104:343--355.

\bibitem{WaK}
Wu MS, Kirchner HOK (2010) Nonlinear elasticity modeling of biogels. 
J. Mech. Phys. Solids 58:300--310.

\bibitem{Spencer}
Spencer AJM  (1984) Constitutive theory for strongly anisotropic solids. In 'Continuum Theory of the Mechanics of Fibre-Reinforced Composites' (A.J.M. Spencer ed.). CISM Courses and Lectures No. 282. Springer-Verlag, Vienna. 
 
\bibitem{me}
Murphy JG (2013)  
Transversely isotropic biological, soft tissue must be modelled using both anisotropic invariants.  
European Journal of Mechanics A/Solids 42:90--96.
 


\bibitem{Dokos}
Dokos S, Smaill BH, Young AA, LeGrice IJ (2002)  
Shear properties of passive ventricular myocardium. 
Am J Physiol Heart Circ Physi 283:H2650--H2659


\bibitem{SaBe02}
Saccomandi G, Beatty MF (2002)
Universal relations for fiber-reinforced elastic materials.
Mathematics and Mechanics of Solids 7: 95--110.

\bibitem{Cism}
Ogden RW (2003) 
Nonlinear elasticity, anisotropy, material stability and residual stresses in soft tissue. 
In Biomechanics of Soft Tissue in Cardiovascular Systems, CISM Courses and Lectures Series no. 441, 65--108, Springer, Wien.

\bibitem{MaO3}
Merodio J, Ogden RW (2005)  
Mechanical response of fiber-reinforced incompressible non-linearly elastic solids. 
International Journal of Non-Linear Mechanics 40:213--227.

\bibitem{Vergo13}
 Vergori L, Destrade M, McGarry P, Ogden RW (2013)
On anisotropic elasticity and questions concerning its Finite Element implementation.
Computational Mechanics 52:1185--1197. 

\bibitem{Genn}
Gennisson J-L, Catheline S, Chaffao S, Fink M (2003) 
Transient elastography in anisotropic medium: application to the measurement of slow and fast shear wave speeds in muscles. 
J. Acoust. Soc. Am. 114:536--541.


\bibitem{Pap}
Papazoglou S, Rump J, Braun J, Sack I (2006) 
Shear wave group velocity inversion in MR Elastography of human skeletal muscle. 
Magnetic Resonance in Medicine 56:489--497.

\bibitem{Sinkus}
Sinkus R, Tanter M, Catheline S, Lorenzen J, Kuhl C, Sondermann E, Fink M (2005) 
Imaging anisotropic and viscous properties of breast tissue by magnetic resonance-elastography. 
Magnetic Resonance in Medicine 53:372--387.



\bibitem{Morrow}
 Morrow DA, Haut Donahue TL,  Odegard GM, Kaufman KR (2010) 
 Transversely isotropic tensile material properties of skeletal muscle tissue. 
 J. Mech. Behav. Biomed. Mater. 3:124--129.


\bibitem{TaN}
Truesdell C,  Noll W (1965)
\emph{The Non-linear Field Theories of Mechanics}
Encyclopedia of Physics (S. Flugge, ed.), Volume III/3. 3$^{rd}$ Edition.
Springer-Verlag, Berlin.

\bibitem{Beatty}
Beatty MF (1989) 
Topics in finite elasticity: Hyperelasticity of rubber, elastomers, and biological tissue. 
Appl. Mech. Rev. 40:1699--1734.


\bibitem{Bayley}
Feng Y, Okamoto RJ, Namani R, Genin GM, Bayly PV (2013) 
Measurements of mechanical anisotropy in brain tissue and implications for transversely isotropic material models of white matter.  
Journal of the Mechanical Behavior of Biomedical Materials 23:117--132.


 \bibitem{HGO}
Holzapfel GA, Gasser TC, Ogden RW  (2000).
A new constitutive framework for arterial wall mechanics and a comparative study of material models. 
Journal of Elasticity 61:1--48.

\bibitem{LeTa94}
Le Tallec P (1994).
Numerical Methods for Nonlinear Three-Dimensional Elasticity.
Handbook of Numerical Analysis, Vol. III. (P.G. Ciarlet and J.L. Lions, Editors)
Elsevier.


\bibitem{adina}
ADINA Theory and Modeling Guide, ADINA R\&D, Inc., Watertown, MA 02472 USA (2005).

\bibitem{manual}
ARES Rheometer Manual (2006)
Rheometrics Series User Manual, Revision J.
TA Instrument--Waters LLC, New Castle, USA.











 \end{thebibliography}
\end{document}